\documentclass[PRB,twocolumn,preprintnumbers,amsmath,amssymb,superscriptaddress]{revtex4}
\usepackage{graphicx}
\usepackage{epstopdf}

 \def \La {$^{139}$La }

\def\etal{{\it et al.}}
\def\T1{$T_1^{-1}$}

\def\TT1{($T_1T)^{-1}$}

\begin{document}

\title{Similar glassy features in the NMR response of pure and disordered La$_{1.88}$Sr$_{0.12}$CuO$_4$}

\author{V. F. Mitrovi{\'c}}
\affiliation{Grenoble High Magnetic Field Laboratory, CNRS, BP 166, 38042 Grenoble Cedex 9, France}

\affiliation{Department of Physics, Brown University, Providence, RI 02912, U.S.A. }

\author{M.-H. Julien$^*$}
\affiliation{Laboratoire de Spectrom\'etrie Physique, UMR5588 CNRS and Universit\'e J. Fourier - Grenoble, 38402 Saint Martin d'H\`{e}res,
France}

\author{C. de Vaulx}
\affiliation{Laboratoire de Spectrom\'etrie Physique, UMR5588 CNRS and Universit\'e J. Fourier - Grenoble, 38402 Saint Martin d'H\`{e}res,
France}

\author{M. Horvati{\'c}}
\affiliation{Grenoble High Magnetic Field Laboratory, CNRS, BP 166, 38042 Grenoble  Cedex 9, France}

\author{C. Berthier}
\affiliation{Grenoble High Magnetic Field Laboratory, CNRS, BP 166, 38042 Grenoble  Cedex 9, France}

\author{T. Suzuki}
\affiliation{Advanced Meson Science Laboratory, Nishina Center, RIKEN, 2-1 Hirosawa, Wako, Saitama 351-0198, Japan}

\author{K. Yamada}
\affiliation{Institute for Materials Research, Tohoku University, Katahira 2-1-1, Sendai, 980-8577, Japan}

\date{\today}

\begin{abstract}

High $T_c$ superconductivity in \mbox{La$_{2-x}$Sr$_{x}$CuO$_4$} coexists with (striped and glassy) magnetic order. Here, we report NMR
measurements of the $^{139}$La spin-lattice relaxation, which displays a stretched-exponential time dependence, in both pure and
disordered $x = 0.12$ single crystals. An analysis in terms of a distribution of relaxation rates $^{139}T_1^{-1}$ indicates that i) the
spin-freezing temperature is spatially inhomogeneous with an onset at $T_g^{\rm onset}=20$~K for the pristine samples, and ii) the width
of the $T_1^{-1}$ distribution in the vicinity of  $T_g^{\rm onset}$ is insensitive to an \mbox{$\sim$1 \%} level of atomic disorder in CuO$_2$
planes. This suggests that the stretched-exponential $^{139}$La relaxation, considered as a manifestation of the systems glassiness, may
not arise from quenched disorder.

\end{abstract}

\maketitle

\section{Introduction}

\begin{figure}[t!]
\vspace{-6cm} \centerline{\includegraphics[width=3.4in]{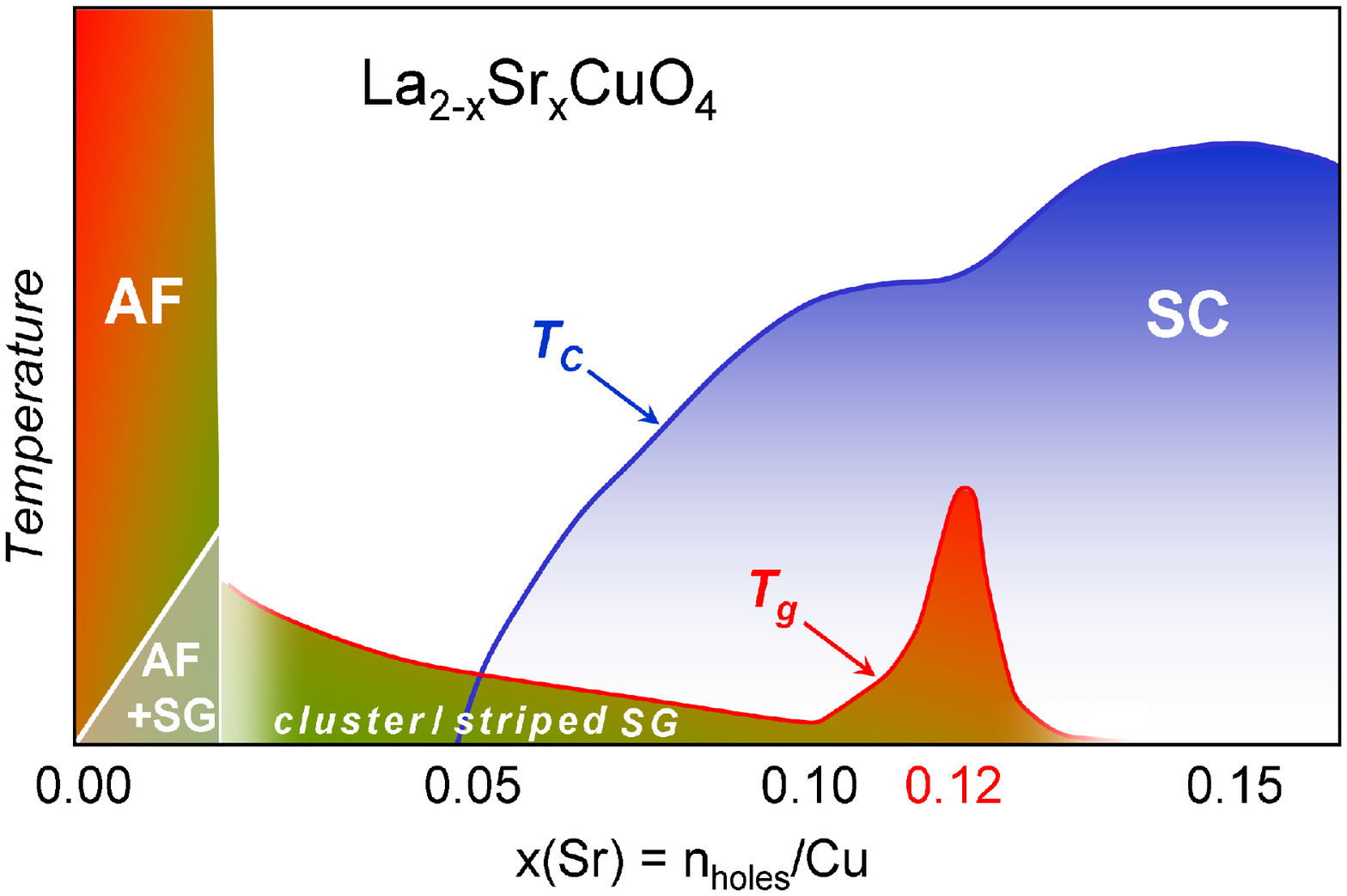}} \caption{\label{phasediag}  (Color online) Sketch of the phase
diagram of La$_{2-x}$Sr$_x$CuO$_4$. AF denotes antiferromagnetic, SC superconducting, and SG  spin glass phase. $T_g$ defines the
temperature of the transition to the frozen state which has been named a ``cluster spin glass" before the discovery of stripes~(see
\cite{Julien03} and Refs. therein). $T_g$ is defined from either NMR or $\mu$SR measurements. These two low energy magnetic probes of
comparable (though not necessarily identical) timescales give similar values of $T_g$, enabling one to obtain a coherent phase
diagram~\cite{Julien03}. Remarkably, $T_g$ (together with the spin stiffness -~see text) is enhanced for $x\simeq 0.12$. One of the
questions regarding the cluster spin-glass phase is whether the non-monotonic behavior of $T_g~vs.~x$ suggests different physics for
$x\simeq 0.12$ than for $x\leq0.10$. Note that the main difference between La$_{2-x}$Sr$_x$CuO$_4$ and compounds presenting direct
evidence for charge stripe order ({\it e.g.} La$_{2-x}$Ba$_x$CuO$_4$) is that the peak of $T_g~vs.~x$ around $x~\simeq~0.12$ is
considerably wider in these latter compounds~\cite{Klauss}.}
\end{figure}

The coexistence of magnetic order with superconductivity is a prominent feature of \mbox{La$_{2-x}$Sr$_x$CuO$_4$ \cite{Julien03}} and
other underdoped high $T_c$ cuprates~\cite{magneticorder,Klauss}, even in zero magnetic field. However, neither the origin of this static
magnetism nor the reason for its glassy character are fully understood. An important school of thought focuses on stripe
physics~\cite{Kivelson00}. Indeed, the temperature of magnetic freezing, $T_g$, in La$_{2-x}$Sr$_x$CuO$_4$ is peaked in the vicinity of
\mbox{$x\simeq\frac{1}{8}$} \mbox{(Fig. \ref{phasediag})}, and neutron scattering studies~\cite{Suzuki98,Kimura99,Wakimoto01} reveal
long-range antiferromagnetic order with the same typical modulation as in the materials presenting direct evidence for charge stripe
order~\cite{Tranquada}. Another approach relies on electronic and magnetic inhomogeneities generated by quenched
disorder~\cite{spinglass}. Undoped or weakly hole-doped droplets may form the magnetic clusters which freeze at low temperature, as
suggested by the recent nuclear magnetic resonance (NMR) evidence for a nanoscale inhomogeneity of the hole concentration in
La$_{2-x}$Sr$_x$CuO$_4$~\cite{Singer}. Furthermore, the importance of quenched disorder could be supported by the glassy features observed
in superconducting samples ($x \geq 0.06$)~\cite{Julien03}. These are reminiscent of the spin-glass behaviour, well documented for
$0.03\leq x\leq0.05$. Spatial heterogeneity is particularly evident in a number of magnetic measurements, such as the
stretched-exponential NMR relaxation of $^{139}$La nuclei. This heterogeneity develops as the magnetic fluctuations slow down over a
substantial temperature range on cooling above the freezing temperature $T_g$, the value of which depends on the timescale of the
measuring probe. These properties are typical of glassy systems.

Glassiness is, however, also present in materials where stripe order is well-defined and relatively long range~\cite{Tranquada}.
Furthermore, from the $T$ dependence of an average $^{139}$La spin-lattice relaxation rate $T_1^{-1}$, Curro~\etal~\cite{Curro} concluded
that the distribution of magnetic correlation times is similar in several materials with very different hole or impurity contents. This
has led to the suggestion that glassiness in these materials is not due to quenched disorder but is self-generated by the charge
stripes~\cite{Schmalian}. Later, Hunt \etal \cite{Hunt01} determined the distribution of correlation times in stripe-ordered materials in
a more direct way than Curro \etal \cite{Curro}, {\it i.e.} from the recovery of the $^{139}$La NMR signal. The stretched-exponential
relaxation was thus better characterized, but its origin was not the focus of the discussion.

In this paper, we report $^{139}$La NMR measurements in La$_{1.88}$Sr$_{0.12}$CuO$_4$, where $T_g$ is maximum, as illustrated in
\mbox{Fig. \ref{phasediag}}. We quantify the $^{139}$La stretched-exponential relaxation and discuss the issues of inhomogeneity and
atomic disorder.

\section{Samples}

Single crystals, of typical size \mbox{$2\times 3 \times 4$~mm$^3$}, were cut from rods grown by the traveling solvent floating zone
method. Growth at a rate of $\simeq$1~mm/h yielded ``pristine'' samples with the standard $T_c$ of 30~K~\cite{Suzuki98}, while growth at
0.2~mm/h produced a ``disordered'' sample with a low $T_c$ of 10~K ($T_c^{\rm onset}=12$~K)~\cite{Katano00}. Neutron scattering (NS)
studies have shown that the two crystals have the same structure as well as an identical temperature of the transition from the high $T$
tetragonal (HTT) phase to the low $T$ orthorhombic (LTO) phase. These studies have also revealed an identical incommensurability of the
magnetic peaks. Since these quantities depend strongly on $x$, the results demonstrate that the two crystals have the same doping
$x=0.12$. On the other hand, magnetic Bragg peaks appear below $T_g^{\rm NS}=30$~K~\cite{Suzuki98} and 25~K~\cite{Katano00} for the
pristine and disordered samples, respectively. All of these properties, as well as an upturn of the zero-field in-plane resistivity below
80~K, are consistent with the presence of $\sim1$~\% of non-magnetic defects in the disordered sample~\cite{Koike92}. These defects likely
correspond to Cu vacancies, which produce the same magnetic effects as non-magnetic impurities~\cite{Julien00,Rullier}.

\begin{figure}[t!]
\vspace{-5cm}
\centerline{\includegraphics[width=8.1cm]{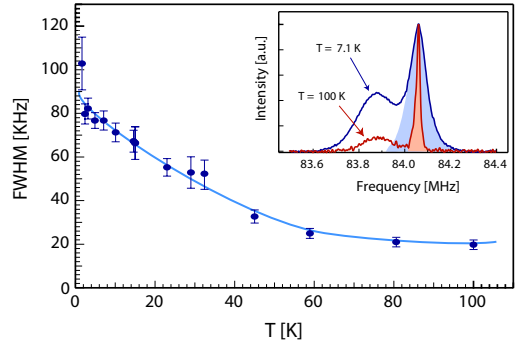}} 
\caption{\label{Spec}   (Color online) The temperature dependence of  \La linewidth (filled  symbols) for the high frequency peak (shaded
region in the inset) of  $\left\langle { + {1 \over 2}\leftrightarrow - {1 \over 2}}\right\rangle$ transition in the magnetic field
applied at $\theta \sim 10 ^{\circ}$ away from   the $c$-axis for La$_{1.88}$Sr$_{0.12}$CuO$_4$ ($T_c= 30$~K sample). The solid line is
guide to the eye. {\bf Inset:} Normalized   spectra at $T=100$~K  and $T=7.1$~K as denoted. The shaded areas mark the high frequency peak
where the rates presented here are determined. The low frequency peak contains the signal from other quadrupolarly split $\left\langle { +
{1 \over 2}\leftrightarrow - {1 \over 2}}\right\rangle$ transition lines \cite{Julien01}.} \vspace*{-0.4cm}
\end{figure}

\section{NMR methods}

As in~\cite{Julien01}, the applied magnetic field was tilted away from the $c$-axis by $\theta \sim 10 ^{\circ}$, in order to obtain a
sharp peak on the high frequency edge of the $^{139}$La central, $  \left\langle { + {1 \over 2}\leftrightarrow - {1 \over
2}}\right\rangle $,  Zeeman transition. $T_1$ was measured on this peak, shown as a shaded part of the sample spectra  in the inset to \mbox{Fig. \ref{Spec}}.
The $T$ dependence of $T_1$ is identical on other peaks in the central transition spectrum. Experiments were performed in fields of 9~T
(for $x=0.12$, $T_c=10$~K sample) and 14~T (for $x=0.12$, $T_c=30$~K, and for $x=0.10$ samples).

In \mbox{Fig. \ref{Spec}}, we plot the temperature dependence of  \La linewidth  for the high frequency peak (shaded region in the inset)
of $\left\langle { + {1 \over 2}\leftrightarrow - {1 \over 2}}\right\rangle$ transition   for La$_{1.88}$Sr$_{0.12}$CuO$_4$ ($T_c= 30$~K
sample). The linewidth broadens with decreasing temperature indicating increasing distribution of local static magnetization as the
temperature is lowered.

We remark that even at the lowest $T$ the signal from other central transition lines is insignificant at the frequency of the the high
frequency peak  and its   high frequency side, for this orientation of the applied field. Therefore, the temperature evolution of the
measured relaxation rates is intrinsic to magnetism and not a result of signal overlap from other central transition lines.

We also point out that no sign of phase separation is detected in our NMR data, in apparent disagreement with $\mu$SR data showing a
magnetic volume fraction of $\sim$20\% for $x=0.12$~\cite{Savici02}. Since NMR was performed here in a high magnetic field, part of the
discrepancy might be resolved by an increase of the magnetic volume fraction with the applied field~\cite{Savici05,Machtoub05}.

\section{Analysis of $T_1$ data}

\begin{figure}[t!]
\vspace{-6cm}
\centerline{\includegraphics[width=3.5in]{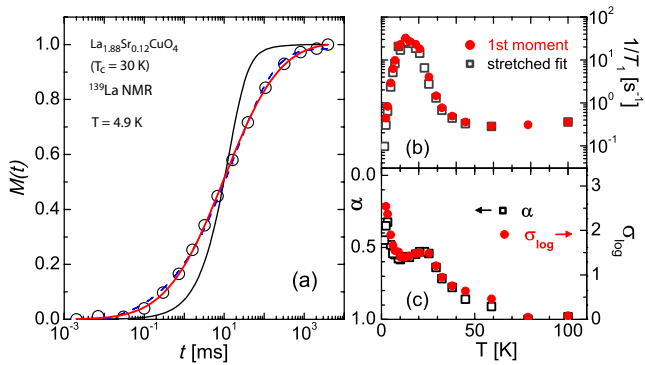}} 
\caption{\label{recovery}   (Color online) {\bf (a)} An example of the  time dependence of the \La normalized longitudinal
magnetization ${\cal M}(t)= M_z(t)/M_z(\infty)$ after saturation of the central transition (dots) in La$_{1.88}$Sr$_{0.12}$CuO$_4$ ($T_c=
30$~K sample). The thin black line is a fit to Eq.~1 with $\alpha=1$. The blue (dark gray) dashed line is a fit to Eq.~1 with
$\alpha=0.48$. The thick red (light gray) line is the fit to \mbox{Eq. \ref{ProbT1}}. {\bf (b)} and  {\bf (c)} Comparisons of results
obtained from the stretched-exponential fit (open symbols) and from the Gaussian distribution fit (filled  symbols). See text for
details.}
\end{figure}

In \mbox{Figure \ref{recovery}}, we display the time ($t$) dependence of the \La longitudinal magnetization ${\cal M}(t) = M_z(t)
/M_z(\infty)$ after a comb of $\pi \over 2$ saturation pulses, in the pristine sample. Clearly, the data cannot be fitted by the
theoretical formula~\cite{Narath67}, which corresponds to $\alpha=1$ (``exponential'' relaxation) in
\begin{eqnarray}
\label{REcovSE}
{\cal M}_{\alpha}(t, T_1^{-1})   &=  & 1 - \,0.714 \, e^{-\left( 28{t \over T_1}\right)^\alpha} - \, 0.206\, e^{-\left(15{t \over
T_1}\right)^\alpha} \nonumber \\
&&- \, 0.068 \, e^{-\left(6{t \over T_1}\right)^\alpha}  - \,0.012\, e^{-\left({t \over T_1}\right)^\alpha}.
\end{eqnarray}
We point out that the above statement is true for data at any temperature below \mbox{$\approx 80$ K}, as evident in \mbox{Fig. \ref{recovery}c} when $\alpha < 1$.
In La$_{2-x}$Sr$_x$CuO$_4$ materials, the stretched relaxation, {\it i.e.} the deviation from ${\cal M}_{\alpha = 1}(t,
T_1^{-1})$, has been attributed to a distribution of $^{139}T_1^{-1}$ values. We shall comment on this interpretation below.

In an ideal case, one-to-one correspondence between dynamic and static inhomogeneities can be revealed by measuring \T1~as a function of
position across the NMR line shape~(see~\cite{Singer,mitrovi01} for example). However, despite the observation of a continuous
line-broadening (Shown in Fig.~2), no significant frequency dependence of $T_1^{-1}$ was found across the $^{139}$La line. Thus, to
quantify the inhomogeneities one must resort to alternative analysis of ${\cal M}(t)$ data, in two possible ways:

\begin{figure}[t!]
\vspace{-3cm} \centerline{\includegraphics[width=8.2cm]{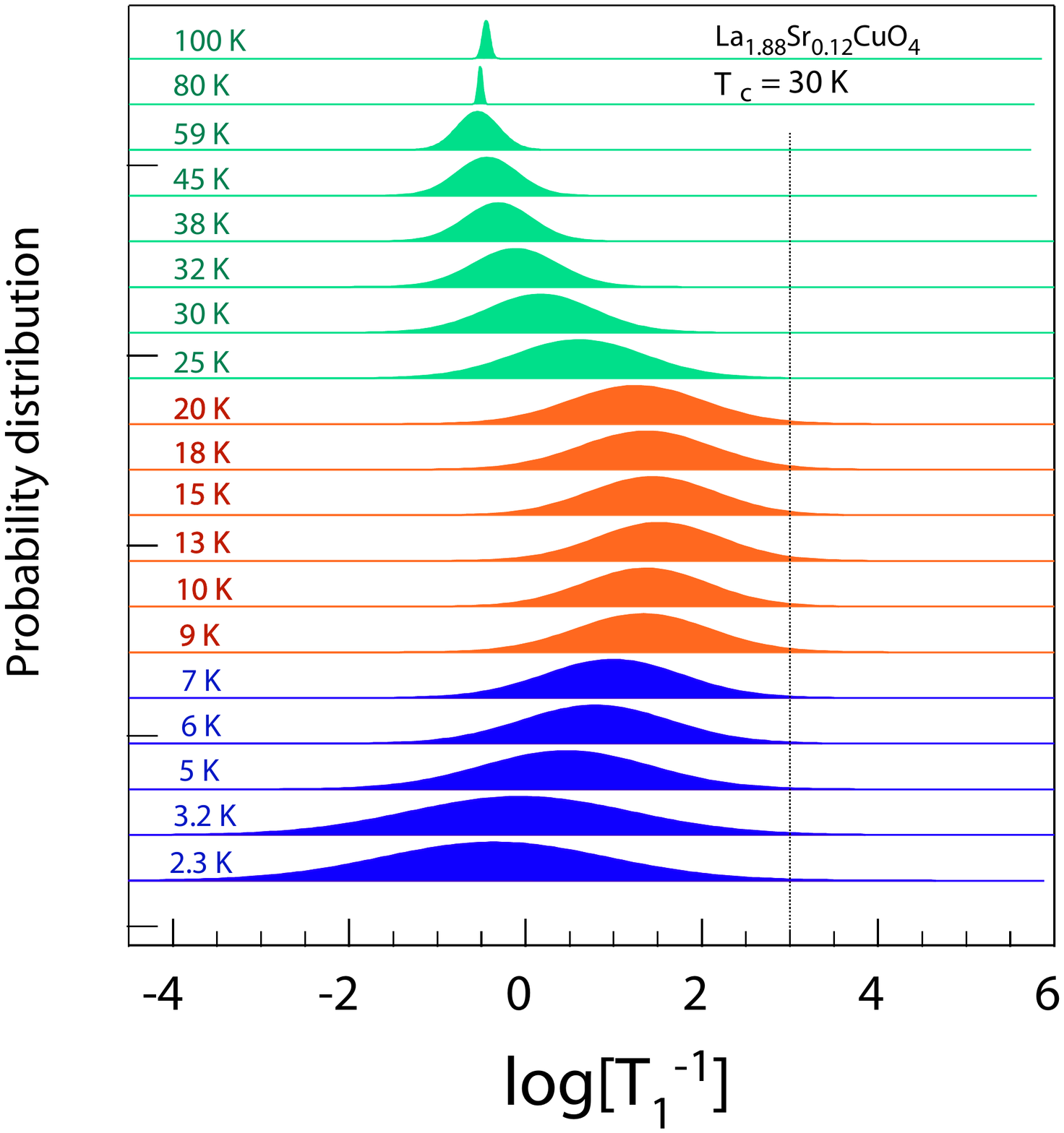}} \vspace{-0.4cm} \caption{\label{distrib30K}  (Color online)
The Gaussian probability distribution of the relaxation rate  as a function of $T$ for La$_{1.88}$Sr$_{0.12}$CuO$_4$ ($T_c= 30$~K sample).
The rates are in units of s$^{-1}$. The  amplitude of distributions are normalized to one and curves offset for clarity. The vertical line
indicates a typical maximum value $(T_1^{-1})_{\rm max}\simeq10^3$~s$^{-1}$ in the system. Green color (lightest gray) depicts the $T$
interval (from 100 K to 25 K) of slowing down, red (lighter gray) represents the $T$ range (from 20 K to 9 K) where spins freeze
throughout the sample, and navy blue (dark gray) is used when all of the sample is magnetically frozen on the NMR timescale.}
\vspace{-0.2cm}
\end{figure}

(i) First, a fit to \mbox{Eq. \ref{REcovSE}} is made with the stretching exponent $\alpha \neq 1$. This provides a  phenomenological
account of the distribution of $T_1$ values (Fig.~\ref{recovery}).

(ii) Second, the formula ${\cal M}_{\alpha = 1}(t, T_1^{-1})$, is convoluted with a chosen probability distribution function of
$T_1^{-1}$. We found that a good fit to the data is provided by the Gaussian distribution on a logarithmic $\left(\textrm{log}\,T_1^{-1}
\right)$ scale:
\begin{eqnarray}
\label{ProbT1} &&  {\cal M}_\textrm{G}(t) = (\sqrt{\pi/2}~ \sigma_{\textrm{log}})^{-1}
\times \qquad \nonumber \\
&&\int \textrm{e}^{-2 \left(\textrm{log}\,R_1 - \textrm{log}\,T_1^{-1} \right)^2/\sigma_{\textrm{log}}^2} ~M_{\alpha = 1}(t, R_1)
~d(\textrm{log}\,R_1). \; \;\;\;\;\;
\end{eqnarray}
This fit is defined by only two parameters: the most probable relaxation rate $T_1^{-1}$, ({\it i.e.}, the center of the Gaussian), and
the width of the distribution $\sigma_{\textrm{log}}$ on a log$_{10}$ scale. Using the $\textrm{log}\,T_1^{-1}$ space is more physical
when very broad distributions of relaxation rates are expected. This also naturally avoids the introduction of an artificial low
$T_1^{-1}$ cut-off needed to eliminate unphysical negative values encountered when the linear scale is used. The fit yields $T_1^{-1}$
values equivalent to the stretched fit $M_{\alpha}(t, T_1^{-1})$, as confirmed from the analysis of our data shown in \mbox{Fig.
\ref{recovery}b}. Furthermore, the value of $\sigma_{\textrm{log}}$ directly shows over how many orders of magnitude spreads the
distribution of relaxation rates. Direct insight into this parameter is the main advantage of this fit (see \mbox{Fig. \ref{distrib30K})}.
The $\sigma_{\textrm{log}}$  also  appears to be linearly related to the value of $\alpha$, as demonstrated  in \mbox{Fig. \ref{recovery}c},  in agreement with
predictions of the recent theoretical work of Johnston for $\alpha\geq 0.5$~\cite{Johnston06}. We also remark that we cannot experimentally determine the
unique/exact shape of the distribution function. Equally good fits of the data can be achieved by assuming, {\it e.g.} an asymmetric Gaussian or
a Lorentzian distribution for \T1. Thus, we choose to analyze the data assuming    a distribution function of the simplest form, a Gaussian.
However, the results to be discussed below have been found to be insensitive to the choice of the exact
form of the distribution.

\begin{figure}[t!]
\vspace{-2.5cm}
\centerline{\includegraphics[width=8.2cm]{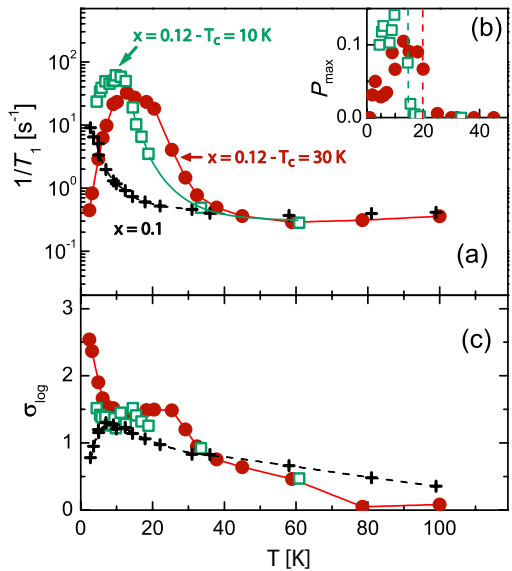}} 
\caption{\label{results}   (Color online) {\bf (a)} $^{139}$La spin-lattice relaxation rate $T_1^{-1}$ defined as the center of the
Gaussian distribution ${\cal P}$, for the two $x=0.12$ samples and for an $x=0.10$ sample.  {\bf (b)} ${\cal P}_{\rm max}={\cal
P}(\log(T_1^{-1})_{\rm max})/{\cal P}(\log(T_1^{-1}))$ where \mbox{$(T_1^{-1})_{\rm max}=10^3$~s$^{-1}$} (same symbol code as in main
panel). Vertical dashed lines denote $T=15$ and 20~K, where the first spins become frozen in the samples. {\bf (c)} Width $\sigma_{\rm
log}$ of ${\cal P}(\log T_1^{-1})$ (same symbol code as in upper panel). The lines are guides to the eye. The downturn of $\sigma_{\rm log}$ at low $T$ for $x=0.10$ might be due to a loss of NMR signal. }
\vspace*{-0.2cm}
\end{figure}

\section{Temperature dependence of the spin dynamics}

In this section, we shall first give a qualitative description of the temperature dependence of the mean relaxation rate $T_1^{-1}$, in
order to define the freezing temperature $T_g$ on the timescale of NMR. Then, tentative fits of the $T$ dependence of $T_1^{-1}$ will be
described. This parametrization of the data enables primarily a comparison between
different La$_{2-x}$Sr$_x$CuO$_4$ samples, while it cannot  providing a
physical picture of the spin-glass freezing in the system.
 Furthermore, the shortcomings of such  analysis
based on the mean value of the distribution of $T_1^{-1}$ values are alluded to  in the next section.

The temperature dependence of $T_1$ may  be understood, at least qualitatively, from the following expression:
\begin{equation}
\label{EqBpp}
{1 \over T_1} = \gamma_n^2 \left< h_{\bot}^2 \right> {2 \tau_c \over
{1 + \omega_n^2 \tau_c^2}},
\end{equation} where $\tau_c$ is the correlation time, $h_\bot=(h_{xx}^2+h_{yy}^2)^{1/2}$
the component of the hyperfine field perpendicular to the field direction,  and $\omega_n$ the NMR frequency~\cite{BPP}. At high
temperatures the correlation time is short, that is the condition $\tau_c^{-1}\gg \omega_n$ is satisfied. As the dynamics of the system
slows down on cooling, $\tau_c^{-1}$ decreases causing an increase of $T_1^{-1}$. This occurs down to the temperature $T=T_g^{\rm NMR}$ at
which the condition $\tau_c^{-1} = \omega_n$ is reached. Upon further cooling, $\tau_c^{-1}$ continues to  decrease but $T_1^{-1}$
decreases. Thus, $T_g^{\rm NMR}$, the temperature of freezing on the timescale of $^{139}$La NMR ($\omega_n\simeq 10 ^{8}$~Hz), is defined
as the temperature at which the relaxation rate is at its maximum value:
\begin{equation}
\left(T_1^{-1} \right)_{\rm max} \equiv T_1^{-1}(T_g^{\rm NMR})=\gamma_n^2 \left< h_{\bot}^2 \right> \omega_n^{-1}.
\end{equation}

For our pristine $x=0.12$ sample, the peak of the mean $T_1^{-1}$ occurs at $T^{\rm NMR}_g = 13$~K. The temperature at which the increase
of $T_1^{-1}$ becomes noticeable may be defined as \mbox{$T^{\rm slow}= 45$ K}. Interestingly, the ratio $T^{\rm slow}/T^{\rm NMR}_g\simeq
3.5 \pm 0.5$ is much lower for $x=0.12$ than for $x=0.10$ (\mbox{$T^{\rm slow}/T^{\rm NMR}_g \gtrsim 7.4$}) and for other values of
$x$~\cite{Cho92,Julien99}. This statement holds even if an onset $T_g^{\rm NMR}$ is considered as described in the next section. Note also
that the ratio is believed to be magnetic field independent for fields ($H_0 \leqslant 14\; {\rm T}$) investigated here \cite{Julien01}.

A more quantitative approach requires an analysis of the temperature dependence of $T_1$. Although it is not clear on which theoretical
model such an analysis should be based, two models which have been used in the context of spin-freezing in the cuprates can be used.

\begin{figure}[t!]
\vspace{-5cm}
\centerline{\includegraphics[width=8.3cm]{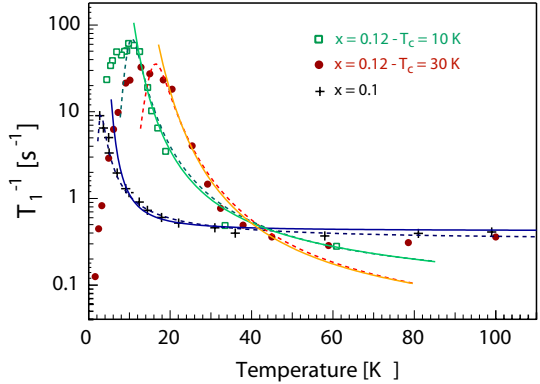}} 
\caption{\label{FitsT1}   (Color online) Fits to the  relaxation rate $T_1^{-1}$ data as plotted in \mbox{Fig.
\ref{results}a}. The dashed lines are fits to the \mbox{Eq. \ref{EqBpp}} with $\tau_c = \tau_0 \exp(E_a/k_BT)$. The solid lines are fits
to the \mbox{Eq. \ref {RCeq}} for both $x=0.12$ samples and to the \mbox{Eq. \ref {RCeqHT}} for $x=0.10$, as described in the text. } \vspace*{-0.2cm}
\end{figure}

First, we use \mbox{Eq. \ref{EqBpp}} with an activated correlation time $\tau_c = \tau_0 \exp(E_a/k_BT)$ to fit the temperature dependence
of $T_1^{-1}$ data with $\tau_0$ and $E_a$ as fitting parameters, as depicted in \mbox{Fig. \ref{FitsT1}}. This fit allows to extract the
effective `energy barrier' $(E_a)$ for the activation of a thermally driven spin freezing process. Fitting the data for $T_g^{\rm NMR}< T
< T^{\rm slow}$, we find the effective energy barrier \mbox{$E_a = 140 \pm 30 \; {\rm K}$} for our pristine $x=0.12$ sample. This value is to be
compared with $E_a = 84 \pm 20 \; {\rm K}$ for the disordered $x=12 \%$ sample and $E_a =  13 \pm 3 \; {\rm K}$ for $x=10 \%$.

Another possibility is to use the renormalized classical form \cite{Chakravarty90} of the 2D Heisenberg model on a square lattice,  which
also captures well the increase of $T_1^{-1}$ with decreasing temperature above $T^{\rm NMR}_g$ (see also~\cite{Hunt01}). This form allows
to extract the effective spin stiffness $(\rho_s)$ parameter. As illustrated in  \mbox{Fig. \ref{FitsT1}},  in the temperature range
$T_g^{\rm NMR} \lesssim   T \lesssim   T^{\rm slow}$, the data for both $x=0.12$ samples fit well to the  low temperature limit \mbox{$(T
\lesssim   2\pi \rho_s/2)$} of the renormalized classical form, expressed as
%
\begin{equation} \label{RCeq} {1 \over T_1} \propto {{e^{C/T}
\cdot (T/C)^{3/2} } \over {C \cdot (1 + T/C)^3}}.
\end{equation}
%
Here \mbox{$C \equiv 2 \pi \rho_s$} is a fitting parameter. We find a value of $\rho_s = 25 \pm 5 \; {\rm K}$ for the spin stiffness of
our pristine $x=0.12$ sample.  For the disordered $x=0.12$ sample, \mbox{$\rho_s = 19 \pm 4 \; {\rm K}$}  is lower. An even weaker value $(\rho_s
= 2 \pm 1 \; {\rm K})$ is found for the $x=0.10$ sample. This correlates with the
fact that \mbox{$T^{\rm slow}/T^{\rm NMR}_g \gtrsim
7.4$} for this sample. Due to its weak spin stiffness parameter, the data for the $x=0.10$ sample are fitted to the high temperature limit
\mbox{$(T  \gtrsim   2\pi \rho_s/2)$}  of the renormalized classical form \cite{Chakravarty90}, given by
\begin{equation}
\label{RCeqHT} {1 \over T_1} \propto
 \left (1 +  \frac{C}{4T} \right )^{1/2}  \cdot
 \exp {\left [ \left (1 + \frac{\small C}{\small 4T} \right )  \left ({C \over {2T}} \right )^2 \right ]},
\end{equation}
with \mbox{$C \equiv 2 \pi \rho_s$} as a fitting parameter. Not surprisingly (given the similarities of the fitting formulas in this
temperature range), the extracted values of the spin stiffness $ 2 \pi \rho_s$ are comparable to the values of the effective energy
barrier in all the three samples. On the other hand, the spin-stiffness values are smaller than those in La-based cuprates with LTT
structure and charge-stripe order~\cite{Hunt01,Teitelbaum03}.

In conclusion, it appears that \mbox{$x=0.12$} is the concentration for which the temperature range of slowing down of the magnetic
fluctuations is the least extended and/or the spin-stiffness is the strongest. We remark that the last point is in agreement with results
in Ref.~\cite{Teitelbaum03}.

\section{Inhomogeneity of $T_g$}

For a more adequate description of the spin freezing, it is in fact the maximum, and not the mean, $T_1^{-1}$ value present in the system
which should be considered. We define this value to be the rate at which the probability distribution is 1\% of its maximum. Based on
results shown in \mbox{Figs. \ref{distrib30K} and \ref{results}b}, we deduce that a typical maximum value is approximately
\mbox{$(T_1^{-1})_{\rm max}\simeq10^3$~s$^{-1}$}. It is apparent in \mbox{Fig. \ref{results}b} that the weight of the Gaussian
distribution at \mbox{$T_1^{-1}=10^3$ s$^{-1}$} becomes sizeable at $T_g^{\rm NMR}(\rm onset)=20$~K. This means that spins begin to be
frozen well above 13~K in some regions of the sample. Our $T_{g,\, {\rm onset}}^{\rm NMR}=20$~K agrees perfectly with the appearance of a
coherent precession of the muon spin in a $x=0.12$ single crystal similar to ours~\cite{Savici02}.

Here, it must be recognized that $(T_1^{-1})_{\rm max}=10^3$~s$^{-1}$ probably also represents a cutoff value above which the $^{139}$La
NMR signal is unobservable (wipeout phenomenon)~\cite{Curro,Hunt01,Julien01}. However, the consistency of the above
analysis~\cite{BPPanalysis} suggests that our data correctly describe the spin dynamics of the system at least down to 20~K.

For the disordered sample, the freezing occurs at a lower temperature: $T_1^{-1}~vs.~T$ shows a peak at $T=10$~K and $T_g^{\rm NMR,
onset}=$~15~K. This is consistent with
  the qualitative analysis discussed in the previous section and with  the decrease of $T_g^{\rm NS}$ from 30~K (pristine)~\cite{Suzuki98,Kimura99,Wakimoto01} to 25~K
(disordered)~\cite{Katano00}, as determined by neutron scattering. Since the elastic NS signal is integrated over a typical energy window
of $\sim$1 meV and the slowing down of the fluctuations occurs on a relatively wide $T$ range, we observe $T_g^{\rm NS}>T_{g, {\rm
onset}}^{\rm NMR}$.

\section{Inhomogeneity of the spin dynamics}

The distribution of $T_1$ values acquires a significant width only at $T \simeq 60$~K and below. The fact that $\sigma_{\rm log}\simeq 0$
at $T=80$~K and 100~K is remarkable. This means that the distribution of $T_1^{-1}$ values seen by $^{63}$Cu or $^{17}$O NMR at these
temperatures, and attributed to a nanoscale inhomogeneity of the hole concentration~\cite{Singer}, is completely absent in the $^{139}$La
$T_1$ data. This contrasting situation is explained by two facts: First, the hyperfine field at $^{139}$La sites results from the coupling
to several electronic sites in two different CuO$_2$ planes. Thus, the hyperfine field is spatially more homogeneous than the electronic
density. Second, the relatively weak amplitude of this (transferred) hyperfine field produces much weaker differences in the $T_1$ values
for $^{139}$La than for $^{63}$Cu or $^{17}$O. Still, the $^{139}T_1$ inhomogeneity, which shows up at lower temperature, appears to be
correlated to inhomogeneities in the static local magnetization, as revealed by the similar temperature dependence of both the
distribution width $\sigma_{\rm log}$  and the static $^{139}$La linewidth shown in \mbox{Fig. \ref{recovery}c} and \mbox{Fig.
\ref{Spec}}, respectively. However, for the reasons given above, no direct correlation between the inhomogeneity of $^{139}T^{-1}$ and the
   resonance frequency across the  $^{139}$La  line could be established.

Since $^{17}$O and $^{63,65}$Cu NMR signals undergo a significant wipeout below $\sim$50~K~\cite{Julien01, Hunt99}, it is difficult to know
whether the inhomogeneity probed in $^{139}$La $T_1$ measurements below $\sim$60~K is distinct from the one probed at higher temperatures
and attributed to spatial variation of the hole concentration.
At any given temperature below $\sim$20~K, the difference in $^{139}T_1^{-1}$ values between $x=0.12$ and $x=0.10$ samples exceeds an order of
magnitude. However, even if the amplitude of the nanoscale doping inhomogeneity
does not produce a significant effects on the $^{139}$La NMR at  $\sim$100~K, the possibility that it   produces  the large $^{139}T_1$
distribution at low $T$ cannot be discarded.
On the other hand, it is possible that magnetic heterogeneity (ubiquitous in glassy systems)
develops prior to the glass transition in La$_{2-x}$Sr$_x$CuO$_4$, in addition to a nanoscale electronic inhomogeneity. In support to this
is the fact that spin freezing and substantial magnetic inhomogeneity are reported in YBa$_2$Cu$_3$O$_{6+x}$~\cite{magneticorder}, while
a  significant nanoscale variation of the hole concentration is apparently absent in this system~\cite{Bobroff}.

As evident in   \mbox{Fig. \ref{results}b},  $T_1$ values are typically distributed over more than one order of magnitude at low $T$, that
is  $\sigma_{\rm log}$ reaches a value of $\sim 1.5$ at $T=T_g^{\rm NMR}$. This width does not depend strongly on the number of holes
since similar values of $\sigma_{\rm log}$ are observed for $x=0.10$~\cite{Julien01} (Fig.~\ref{results}) and $x=0.06$~\cite{Julien99}
(not shown) at $T\simeq T_g^{\rm NMR}$.

Strikingly, the same value of $\sigma_{\rm log} \simeq 1.5$ is found at $T = T^{\rm NMR}_g$ in both pristine and disordered samples for
$x=0.12$~\mbox{(Fig. \ref{results}c)}. Thus, the distribution width appears to be insensitive to an $\sim 1$\% of in-plane disorder as
well. One could argue that this result can be explained by the fact that La$_{1.88}$Sr$_{0.12}$CuO$_4$ is already a significantly
disordered material in its pristine version. However, this argument does not hold since a typical 1~\% of non-magnetic impurities or
vacancies is highly detrimental to $T_c$ and it clearly affects other magnetic properties as well~\cite{Koike92}.

Finally, we will comment on  alternative explanations of the stretched-exponential behavior of the NMR relaxation. It was suggested that
the inhomogeneous magnetic state is related to extended charge density waves with imaginary order parameter \mbox{($id$-CDW)}
\cite{Rigamonti}. Within this model, the observed inhomogeneity in the NMR and $\mu$SR quantities would originate from  sliding motions of
orbital currents coexisting with the $d$-wave superconducting state. It is known that  any order parameter of  $d$-wave symmetry should be
highly sensitive to impurities. This is in contrast  to  our data,  showing that the inhomogeneity of $^{139}T_1$ is insensitive to 1\% of
atomic disorder. Therefore, the \mbox{$id$-CDW} scenario  is unlikely to account for  our observations.

Furthermore, a power-law time dependence of the spin-spin correlation function, instead of a distribution, is also debated in canonical
spin-glasses  as an explanation for the stretched-exponential NMR or $\mu$SR relaxation~\cite{correlation}.  Anyhow, we did not consider this
possibility as the presence of magnetic heterogeneity in La$_{2-x}$Sr$_x$CuO$_4$ is established     by various experimental facts,
such as the partial wipeout of the NMR signal or the $T$-dependent broadening of the NMR linewidth.

\section{Conclusion}

We have presented an NMR investigation of La$_{1.88}$Sr$_{0.12}$CuO$_4$, a prototypal material for studying spin-glass and stripe-ordering
instabilities in the superconducting regime. Our analysis can be viewed as a parametrization of the stretched-exponential spin-lattice
relaxation of $^{139}$La nuclei. We observe that this phenomenon is not affected by a $\sim$1~\% level of disorder. This result might thus
support proposals~\cite{Schmalian} of heterogeneous dynamics, or more generally of glassiness, which is not due to quenched disorder.
However, how much of the magnetic heterogeneity/glassiness of La$_{2-x}$Sr$_x$CuO$_4$ is attributable to nanoscale variation of the hole
doping remains unclear. It would also be interesting to investigate whether our results for the special $x=0.12$ material still hold in
samples with lower Sr concentration, including the non superconducting cluster spin-glass region of the phase diagram. The glassy nature
of magnetism in superconducting cuprates clearly calls for further theoretical and experimental consideration, especially given the recent
observation of an electronic glass by scanning tunneling microscopy~\cite{Kohsaka07}.\\

\section{Acknowledgments}

We acknowledge useful communications with Roberto de Renzi and Jeff Sonier.

\vspace{-0.5cm}
\bibliographystyle{unsrt}

\begin{thebibliography}{10}
\bibitem{slowing} The real onset of slowing down might occur at a somewhat higher temperature , if it is masked by a $T$
dependent quadrupolar contribution to $^{139}T_1$.
\end{thebibliography}

\end{document}